\documentclass[aps, prx, twocolumn, ...]{revtex4-2}
\usepackage{graphicx}
\usepackage{amsmath}

\begin{document}
\title{Temporal Coarse-Graining as the Origin of Macroscopic Friction in Quantum Spin Chains via Data-Driven Liouvillian Extraction}
\author{Seiki Saito}
\email{saitos@yz.yamagata-u.ac.jp}
\affiliation{Graduate School of Science and Engineering, Yamagata University, Yonezawa, Japan}
\date{\today}

\begin{abstract}
Understanding the emergence of macroscopic irreversible hydrodynamics from the reversible unitary dynamics of isolated quantum many-body systems remains a fundamental challenge. Conventional approaches often force spin density dynamics into purely diffusive models, obscuring the microscopic interplay of pressure, spin current, and local friction. Furthermore, reconciling true irreversibility with strictly unitary evolution raises fundamental questions about the role of the observer's temporal resolution. In this paper, we introduce a fully data-driven framework based on generalized Extended Dynamic Mode Decomposition (gEDMD) integrated with the Mori-Zwanzig projection. By expanding the observable dictionary to explicitly include spin currents, we directly extract the Navier-Stokes hydrodynamic coefficients from a chaotic XXZ spin chain across varying temporal coarse-graining scales. Our unconstrained extraction reveals a clear physical dichotomy: the mechanical elasticity ($c^2$) is intrinsically derived from the exact unitary dynamics, preserving strict microscopic reversibility. In contrast, the macroscopic friction ($\gamma$) and kinematic viscosity ($\nu$) exhibit zero net dissipation, oscillating rapidly around zero in the exact-derivative limit. We demonstrate that genuine macroscopic transport cannot be established without finite temporal coarse-graining. By introducing a finite observation timescale ($\Delta t_{\rm cg} > 0$), the system passes through a distinct crossover timescale where these reversible fluctuations average out, establishing an intermediate functional regime that yields strictly positive friction and viscosity. We further show that the coarse-grained generator constitutes a predictive reduced model requiring only $O(N)$ macroscopic observables, and that the sign of the emergent friction is inherited from the causal, forward-in-time direction of the coarse-grained inference: time-symmetric estimators yield zero net dissipation at any coarse-graining scale. Our results demonstrate that macroscopic friction in isolated quantum systems is not an absolute property, but an emergent phenomenon dictated by the temporal resolution and the causal direction of the observer's coarse-grained description.
\end{abstract}

\maketitle

\section{Introduction}
Understanding how macroscopic irreversible hydrodynamics emerges from the microscopic, reversible unitary dynamics of isolated quantum many-body systems remains one of the central challenges in modern non-equilibrium statistical mechanics \cite{Polkovnikov2011, Gogolin2016}. Under the Eigenstate Thermalization Hypothesis (ETH) \cite{Deutsch1991, Srednicki1994, Rigol2008}, local subsystems are expected to thermalize by acting as a heat bath for one another. Consequently, conserved quantities such as spin density or energy undergo slow macroscopic relaxation, typically described by phenomenological equations like the diffusion equation or the Navier-Stokes equations \cite{Mukerjee2006, Bertini2021}.

In the context of quantum spin chains, the conventional approach to identifying emergent hydrodynamics has heavily relied on fitting the time evolution of the spin density $Z$ directly to Fick's law, thereby extracting a macroscopic diffusion coefficient $D_Z$ \cite{Ljubotina2017, Levi2016}. However, this direct-fitting paradigm possesses fundamental limitations. By assuming an \textit{a priori} purely diffusive behavior ($\dot{Z} = D_Z \nabla^2 Z$), it bypasses the underlying microscopic interplay between the density gradient (pressure), the resulting spin current $J$, and the local friction $\gamma$ induced by the surrounding many-body correlations.

Beyond the spatial constraints of finite-size systems, a more fundamental difficulty lies in the time domain: exact unitary evolution is strictly reversible, meaning that continuous, exact time derivatives cannot inherently generate true macroscopic friction. Thus, reconciling the emergence of dissipative fluid dynamics with the underlying Schrödinger equation requires a careful examination of the observer's temporal resolution, specifically the role of temporal coarse-graining. The questions of \textit{how} macroscopic friction physically emerges and \textit{when} the fluid picture breaks down have remained elusive, largely due to the lack of tools capable of dissecting the dynamics across varying timescales without imposing artificial thermodynamic limits.

In this paper, we overcome these limitations by proposing a fully data-driven approach to systematically investigate the origin of hydrodynamic coefficients from the exact unitary dynamics of a finite chaotic XXZ spin chain. By employing generalized Extended Dynamic Mode Decomposition (gEDMD) \cite{Williams2015, Klus2018, Klus2020} integrated with the Mori-Zwanzig projection formalism \cite{Mori1965, Zwanzig1973}, we systematically expand the observable dictionary to explicitly include both the local spin density $Z$ and the spin current $J$, treating higher-order multi-spin correlations as an environmental heat bath.

Our unconstrained extraction framework allows us to independently evaluate the elasticity $c^2$, the local friction $\gamma$, and the kinematic viscosity $\nu$ across varying temporal coarse-graining scales ($\Delta t_{\rm cg}$). We find a clear physical dichotomy: while the mechanical pressure gradient ($c^2$) is intrinsically derived from the exact time derivative, preserving strict microscopic reversibility, the macroscopic friction ($\gamma$) and viscosity ($\nu$) exhibit rapid, reversible fluctuations around zero with no net dissipation in the exact-derivative limit. We demonstrate that genuine irreversible hydrodynamics, characterized by stable, strictly positive $\gamma > 0$ and $\nu > 0$, only emerges when a finite temporal coarse-graining is introduced. This temporal blurring averages out the microscopic coherent oscillations, driving the system through a distinct crossover timescale into an intermediate functional regime. This emergent fluid picture is, however, inherently transient; it inevitably breaks down either \textit{temporally}, when the coarse-graining scale excessively blurs the macroscopic wave period (the sinc-filter effect), or \textit{spatially}, when finite-size quantum echoes invade the bulk. We further quantify the predictive capability of the extracted coarse-grained generator and its dependence on the size of the observable dictionary, and we show in Appendix~\ref{app:scheme} that the emergent friction is inherited from the causal, forward-in-time direction of the coarse-grained inference: time-symmetric estimators yield zero net dissipation at any coarse-graining scale. Our results demonstrate that macroscopic friction in isolated quantum systems is not an absolute property, but an emergent phenomenon dictated by the temporal resolution and the causal direction of the observer's coarse-grained description.

A movie showing the full temporal evolution of the extracted macroscopic dynamics is available as Supplementary Material at \url{https://doi.org/10.5281/zenodo.20059728}.

\section{Methodology and Simulation System}

\begin{figure}[htbp]
\centering
\includegraphics[width=\columnwidth]{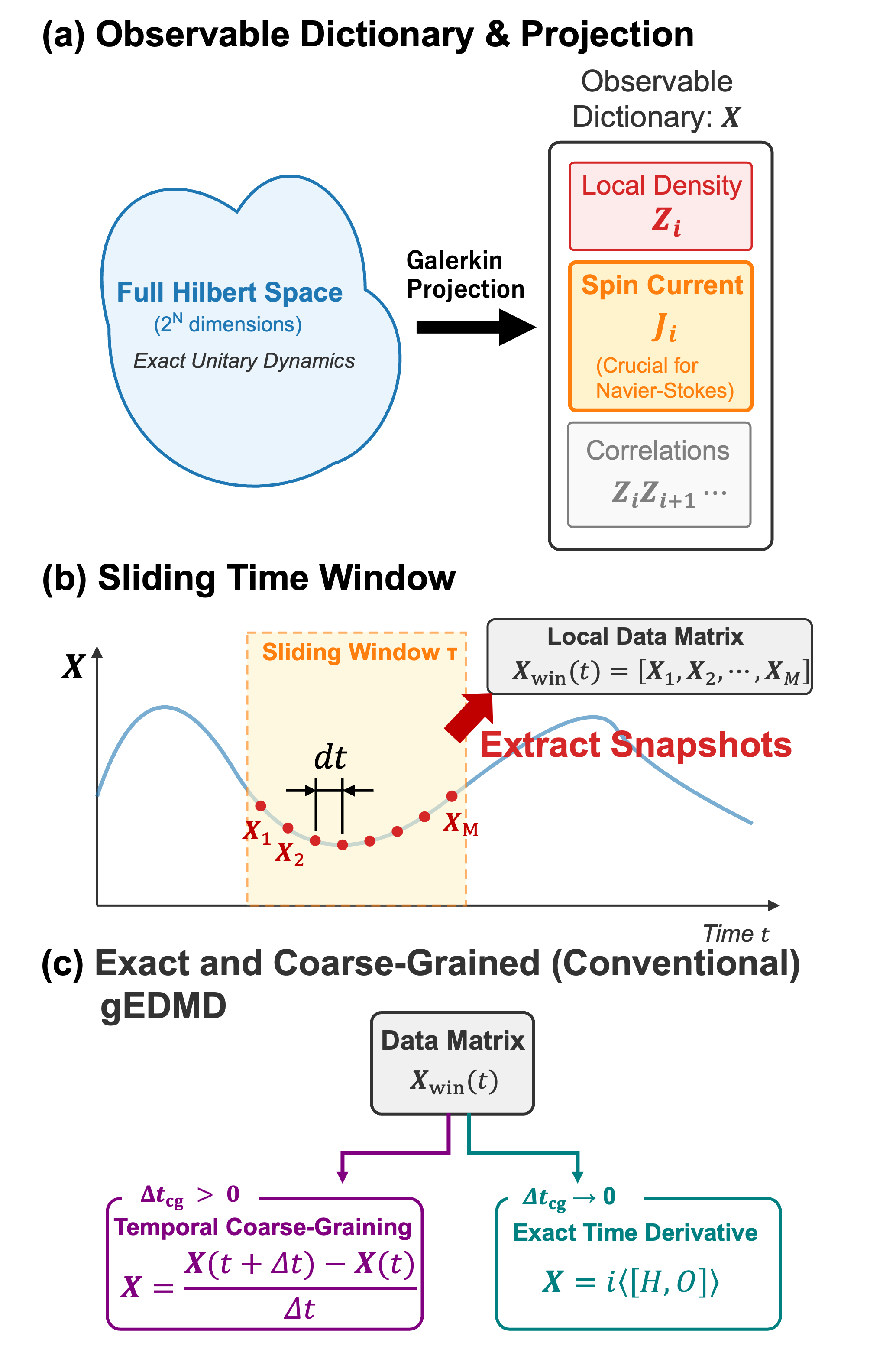}
\caption{\textbf{Data-driven extraction framework and the physical dichotomy of temporal coarse-graining.} (a) Observable dictionary explicitly incorporating both the local spin density $Z_i$ and the spin current $J_i$ as relevant variables extracted from the full Hilbert space via Galerkin projection. (b) Construction of the localized data matrix $\boldsymbol{X}_{\text{win}}(t)$ by extracting discrete snapshots at intervals $dt$ over a sliding time window $\tau$. (c) The two analytical paradigms of the gEDMD framework. The localized data matrix is processed using either an exact continuous time derivative ($\Delta t_{\rm cg} \to 0$) to preserve strict microscopic reversibility, or a finite-difference temporal coarse-graining ($\Delta t_{\rm cg} > 0$) to capture the emergence of macroscopic dissipation.}
\label{fig:schematic}
\end{figure}

\subsection{Quantum Spin Chain Model and Numerical Setup}
To investigate the emergence of macroscopic hydrodynamics from microscopic unitary dynamics, we consider an isolated one-dimensional chaotic XXZ spin-1/2 chain under open boundary conditions. The Hamiltonian is defined as $H = \sum_{i=1}^{N-1} h_{i,i+1} + \sum_{i=1}^{N-2} h_{i,i+2}$, where the nearest-neighbor (NN) and next-nearest-neighbor (NNN) interactions are given by:
\begin{align}
    h_{i,i+1} &= J (X_i X_{i+1} + Y_i Y_{i+1}) + \Delta Z_i Z_{i+1}, \\
    h_{i,i+2} &= J_2 Z_i Z_{i+2}.
\end{align}
Here, $X_i, Y_i, Z_i$ are the Pauli matrices at site $i$. We set $J=1.0$, $\Delta=1.0$, and introduce the NNN interaction $J_2=0.5$ to break integrability and ensure quantum chaotic behavior, which is essential for thermalization \cite{Srednicki1994, Bertini2021}.

The exact unitary time evolution is computed using the state vector $|\psi(t)\rangle$ with a time step of $dt$. To ensure the purity of the unitary evolution without numerical diffusion or the information loss inherent in tensor-network approximations \cite{Ljubotina2017}, the action of the time-evolution operator $|\psi(t+dt)\rangle = e^{-iHdt}|\psi(t)\rangle$ is implemented strictly by employing the Al-Mohy-Higham algorithm \cite{AlMohy2011}. The initial state $|\psi(0)\rangle = \sum_n c_n |n\rangle$ is prepared as a Haar-random-like state in the full Hilbert space: the real and imaginary parts of each coefficient $c_n$ in the computational basis are drawn independently from a uniform distribution with zero mean, and the vector is subsequently normalized. By quantum typicality, the local expectation values of such a state reproduce those of the infinite-temperature (maximally mixed) ensemble up to fluctuations of order $2^{-N/2}$; in particular, the mean energy $\langle \psi(0)|H|\psi(0)\rangle \simeq \mathrm{Tr}\,H/2^N = 0$ and the local magnetization $\langle Z_i \rangle \simeq 0$, ensuring a completely unpolarized ensemble that faithfully captures unbiased thermal fluctuations \cite{Gogolin2016}.

\subsection{Direct Liouvillian Identification via gEDMD and Galerkin Projection}
When extracting governing equations from data, standard Extended Dynamic Mode Decomposition (EDMD) identifies the discrete-time transition operator $K$ from $\boldsymbol{X}(t+\Delta t) = K \boldsymbol{X}(t)$ and reverse-engineers the Liouvillian via the matrix logarithm $L = \frac{1}{\Delta t} \ln K$. However, the matrix logarithm is a multi-valued function, which frequently causes mathematical ambiguities and numerical instabilities when evaluating complex eigenvalues (dissipative poles).

To fundamentally circumvent this, our exact-derivative generator EDMD (gEDMD) approach \cite{Klus2020} bypasses the discrete-time operator $K$ and its matrix logarithm entirely. Let $\boldsymbol{X}(t) = \langle \psi(t) | \boldsymbol{O} | \psi(t) \rangle$ be an $N$-dimensional vector of expectation values corresponding to a chosen dictionary of observables $\boldsymbol{O}$. As illustrated in Fig.~\ref{fig:schematic}(c), instead of utilizing conventional finite differences for temporal coarse-graining ($\Delta t_{\rm cg} > 0$), our framework can directly utilize the exact time derivative computed via the commutator ($\Delta t_{\rm cg} \to 0$), $\dot{\boldsymbol{X}}(t) = i \langle \psi(t) | [H, \boldsymbol{O}] | \psi(t) \rangle$, to identify the finite-dimensional matrix representation $L$ as the generating operator:

\begin{equation}
\dot{\boldsymbol{X}} \approx L \boldsymbol{X}.
\end{equation}
Rather than seeking the true infinite-dimensional operator, deliberately extracting a finite $N$-dimensional matrix $L$, which corresponds to the choice of which physical bases to include in $\boldsymbol{X}$, fundamentally determines the coarse-graining level of the system.

When a finite temporal resolution is considered instead, the time derivative is replaced throughout this work by the causal forward difference, $\dot{\boldsymbol{X}}(t) \approx [\boldsymbol{X}(t+\Delta t_{\rm cg}) - \boldsymbol{X}(t)]/\Delta t_{\rm cg}$, regressed against $\boldsymbol{X}(t)$. This is the first-order estimator of the generator employed by an observer who predicts the future state from the present one. This choice is not innocuous: as analyzed in Appendix~\ref{app:scheme}, the backward difference yields the anti-dissipative mirror image of the forward-difference result, while the time-symmetric central difference reproduces the exact-derivative limit at any $\Delta t_{\rm cg}$. The emergent dissipation reported below is therefore tied to the causal direction of the coarse-grained inference.

For a dataset of $M$ snapshots, gEDMD determines $L$ by minimizing the mean squared error $J(L) = \frac{1}{M} \sum_{m=1}^M \|\dot{\boldsymbol{X}}_m - L\boldsymbol{X}_m\|^2$. Setting $\partial J / \partial L = 0$ yields the solution:
\begin{equation}
L = A G^{-1},
\end{equation}
where $G = \langle \boldsymbol{X}, \boldsymbol{X}^\dagger \rangle$ is the covariance matrix of the bases and $A = \langle \dot{\boldsymbol{X}}, \boldsymbol{X}^\dagger \rangle$ is the cross-covariance matrix with the time derivatives. This least-squares regression is mathematically equivalent to a Galerkin projection onto the low-dimensional subspace spanned by the chosen basis. By requiring the residual $R(L) = \dot{\boldsymbol{X}} - L\boldsymbol{X}$ to be orthogonal to the test functions $\boldsymbol{X}^\dagger$, we obtain the Galerkin condition $\langle R(L), \boldsymbol{X}^\dagger \rangle = 0$, which immediately expands to $\langle \dot{\boldsymbol{X}}, \boldsymbol{X}^\dagger \rangle = L \langle \boldsymbol{X}, \boldsymbol{X}^\dagger \rangle$, perfectly matching the gEDMD formulation.

It is crucial to physically distinguish this data-driven Galerkin projection from the exact Mori-Zwanzig (MZ) projection formalism \cite{Mori1965, Zwanzig1973}. While the structural operation of gEDMD mirrors the MZ projection $\mathcal{P}$ onto the relevant variables, their treatment of the orthogonal complement (the unobserved $\mathcal{Q}$-space) fundamentally differs. The exact MZ equation retains the unobserved dynamics as a non-Markovian memory kernel. In contrast, the data-driven Galerkin projection forcibly truncates this residual at a given instant, demanding a time-local, Markovian generator $L$. Consequently, purely spatial projection without appropriate temporal resolution leads to the immediate breakdown of predictions due to the neglected memory effects (as will be demonstrated in Results I). To physically recover these unresolved memory effects as steady macroscopic dissipation (such as friction and viscosity), the spatial Galerkin projection must be inextricably coupled with temporal coarse-graining, which we introduce in the following section.

\subsection{Dictionary Expansion and Automatic Emergence of Dissipation}
The core strength of our methodology lies in the deliberate restriction of the observation dictionary $\boldsymbol{X}$ to a relatively small number of Hermitian bases. To track macroscopic transport, we systematically expand the dictionary to explicitly include the local 1-body observables (such as spin density $Z_i$ and spin current $J_i = X_i Y_{i+1} - Y_i X_{i+1}$) and 2-body/3-body correlations [Fig.~\ref{fig:schematic}(a)]. 

For instance, in a 30-spin system, the true microscopic state dimension scales to $2^{30}$ (over a billion dimensions); however, our constructed macroscopic dictionary remains confined to merely a few thousand dimensions. During the calculation of the inverse covariance matrix $G^{-1}$ under this drastic dimensionality reduction, the information of the reversible unitary evolution that escapes into the orthogonal complement of unobserved higher-order many-body entanglement is mathematically and automatically converted into non-Hermitian components (eigenvalues with negative real parts) in $L$. Thus, without invoking any \textit{ad hoc} human assumptions such as molecular chaos, the linear algebraic operation of Galerkin projection mathematically converts the unobserved unitary evolution into non-Hermitian components. This spatial dimensionality reduction provides the indispensable mathematical foundation for irreversible dissipative terms, such as macroscopic friction and viscosity, which, as we will demonstrate, fully manifest only when combined with appropriate temporal coarse-graining.

To practically evaluate this temporal coarse-graining and capture the time-resolved dynamics, we employ a sliding time window approach [Fig.~\ref{fig:schematic}(b)]. Rather than computing a single global operator, we construct a localized data matrix $\boldsymbol{X}_{\rm win}(t)$ by horizontally concatenating the snapshot vectors of the observation dictionary across all discrete time steps within a specific window. Specifically, the window duration is set to $\tau = 1.0$, which for the 20-qubit system consists of 500 consecutive data points integrated with an exact time step of $dt = 0.002$. By performing the gEDMD regression directly on this aggregated data matrix $\boldsymbol{X}_{\rm win}$, we ensure that all local temporal fluctuations within the window are comprehensively fed into the Galerkin projection. This allows us to robustly identify the effective time-local Liouvillian $L(t)$ that best approximates the macroscopic dynamics during that specific period.

To analyze this emergence across different scales, we tailor the composition of the dictionary for each specific investigation: we utilize a complete basis for exact validation in small systems (Results I), a partitioned macroscopic-environment basis to directly trace decoherence and information flow (Results II), and a generalized observable basis including length-3 composite correlations to extract the Navier-Stokes hydrodynamic coefficients (Results III). The explicit inclusion of both $Z$ and $J$ allows $L$ to resolve the underlying two-step mechanical structure: $\dot{J} = -c^2 \nabla Z - \gamma J + \nu \nabla^2 J$, granting independent access to the elasticity ($c^2$), friction ($\gamma$), and viscosity ($\nu$).

Finally, rather than relying on artificial superoperator manipulations to mitigate finite-sample statistical noise, we achieve high-precision extraction by scaling up our dataset. By evaluating our generalized gEDMD framework under both the exact and coarse-grained paradigms [Fig.~\ref{fig:schematic}(c)] over a robust ensemble of 500 independent random initial states, horizontally concatenating their time-series data into the aforementioned $\boldsymbol{X}_{\rm win}$, we naturally average out stochastic fluctuations. This large-scale ensemble approach ensures that the extracted hydrodynamic coefficients are statistically converged, avoiding bias in the identified Liouvillian.

\section{Results I: Validation in a Fully Observable System}

Before analyzing the emergence of macroscopic hydrodynamics in large-scale systems, we first validate the fidelity of our Exact-Derivative gEDMD framework. We utilize an 8-qubit chaotic XXZ spin chain, a size small enough to allow for the construction of a complete, exact observable dictionary consisting of all $4^8-1 = 65,535$ Pauli strings.
This fully observable system serves as a controlled testbed to verify that our data-driven approach introduces no algorithmic artifacts and respects the underlying unitary dynamics.

\begin{figure}[htbp]
    \centering
    \includegraphics[width=\columnwidth]{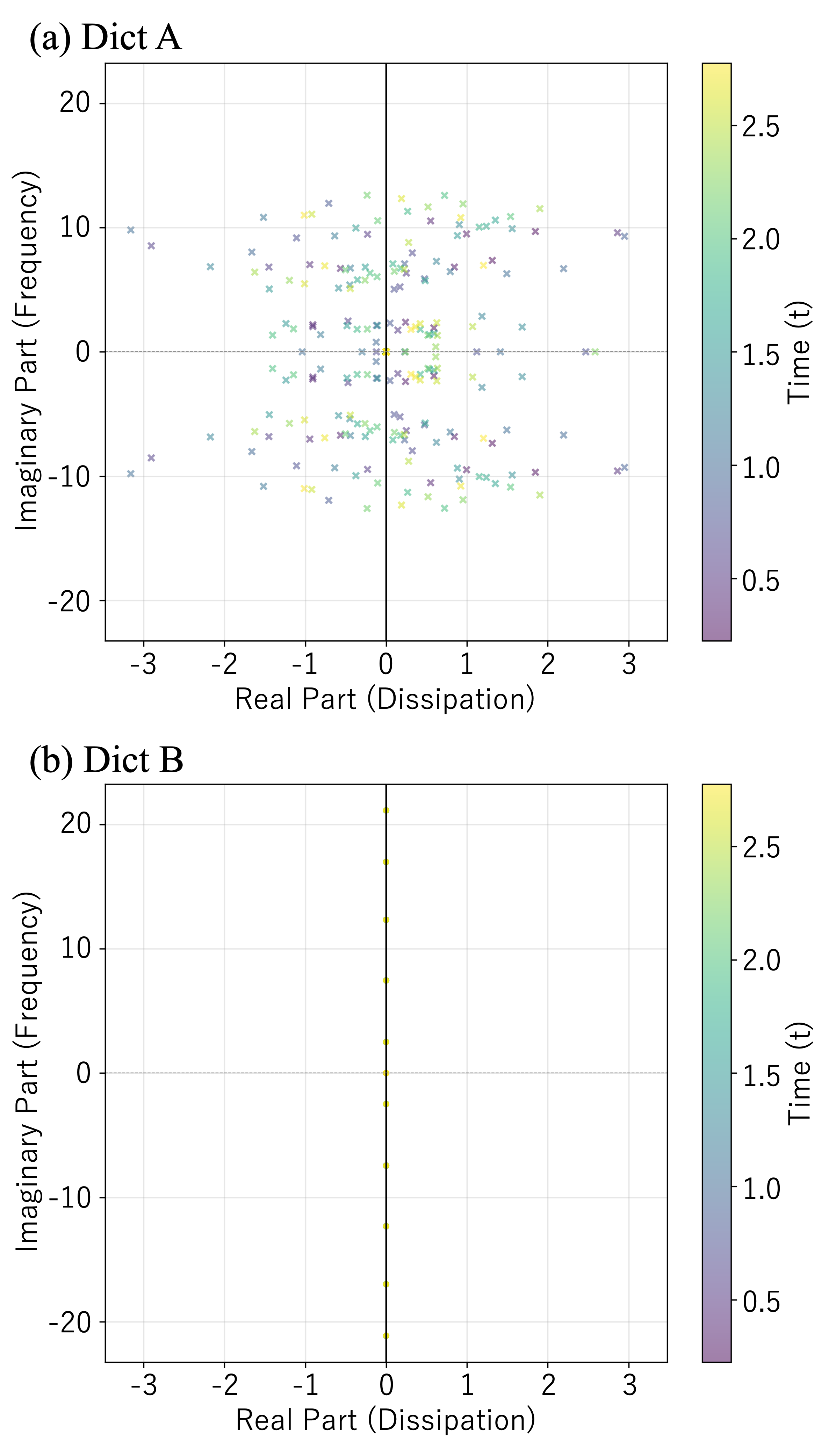}
    \caption{\textbf{Liouvillian spectrum dependence on dictionary completeness.} (a) When using a restricted macroscopic dictionary (Dict A), eigenvalues scatter across the complex plane ($\text{Re}(\lambda) \neq 0$). The emergence of negative real parts reflects apparent dissipation due to information leaking into unobserved microscopic degrees of freedom, while positive real parts indicate apparent amplification driven by finite-size quantum echoes. (b) When utilizing the complete Pauli basis (Dict B), all eigenvalues align on the imaginary axis ($\text{Re}(\lambda) = 0$), recovering the strict reversibility of the isolated quantum system. (An animated version of this spectral evolution is available as Supplementary Video 1.)}
    \label{fig:q8_complex}
\end{figure}

To demonstrate the impact of dictionary selection on the perceived dynamics, we compare the Liouvillian spectrum extracted using a restricted macroscopic dictionary (Dict A) against the full complete dictionary (Dict B). For our $N=8$ qubit system, Dict A comprises $15$ macroscopic observables: the local spin densities $Z_i$ and the nearest-neighbor spatial correlations $Z_i Z_{i+1}$. In contrast, Dict B spans the entire operator space, encompassing all $4^8 - 1 = 65,535$ possible non-identity Pauli strings.
As shown in Fig.~\ref{fig:q8_complex}(a), relying solely on macroscopic observables causes the eigenvalues to spread across the complex plane, dynamically alternating between the left-half ($\mathrm{Re}(\lambda) < 0$) and right-half ($\mathrm{Re}(\lambda) > 0$) planes.
This oscillation signifies a bidirectional flow of information between the macroscopic observables and the orthogonal subspace of unobserved higher-order correlations. Physically, the negative real parts represent apparent dissipation as propagating information generates many-body entanglement that leaks into this unobserved subspace.
Conversely, the positive real parts capture apparent amplification; due to the finite system size, this propagating information hits the spatial boundaries, forcing the quantum echoes to return from the higher-order correlations back into the macroscopic observables.
In contrast, when the complete basis is employed [Fig.~\ref{fig:q8_complex}(b)], the extracted eigenvalues are strictly confined to the imaginary axis ($\text{Re}(\lambda) = 0$).
This confirms that when no information is traced out, our framework captures the reversibility of the isolated quantum system without introducing any artificial damping.

\begin{figure}[htbp]
    \centering
    \includegraphics[width=\columnwidth]{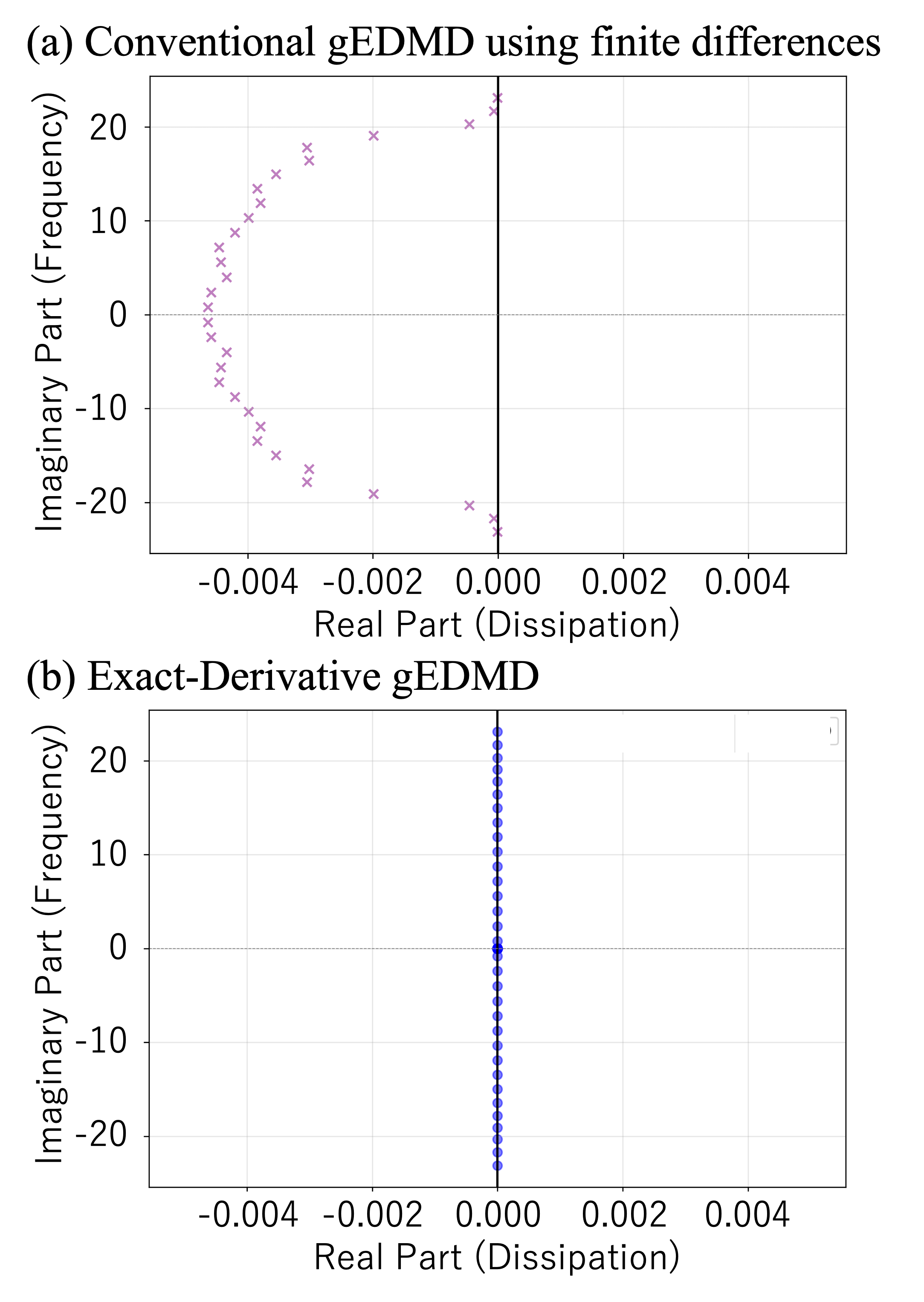}
    \caption{\textbf{The physical dichotomy of temporal resolution.} (a) Applying finite time differences ($\Delta t_{\rm cg} > 0$) causes the eigenvalues to spread into the dissipative left-half plane, illustrating the physical emergence of Markovian dissipation through temporal coarse-graining. (b) Our Exact-Derivative gEDMD preserves the unitary character of the isolated system, ensuring the spectrum remains strictly reversible ($\text{Re}(\lambda) = 0$), a necessary limit for deriving the mechanical elasticity ($c^2$).}
    \label{fig:q8_comparison}
\end{figure}

Furthermore, this fully observable setting highlights the physical dichotomy introduced by the observer's temporal resolution. Figure~\ref{fig:q8_comparison} compares the spectrum obtained via finite-difference temporal coarse-graining ($\Delta t_{\rm cg} > 0$) with our exact-derivative approach. Our exact-derivative gEDMD [Fig.~\ref{fig:q8_comparison}(b)] preserves the unitary character of the isolated system, yielding a strictly reversible, purely oscillatory spectrum. This limit is essential for extracting static, reversible macroscopic quantities such as the mechanical pressure gradient ($c^2$). Conversely, the finite-difference method [Fig.~\ref{fig:q8_comparison}(a)] causes the eigenvalues to spread into the dissipative left-half plane. Rather than being a mere numerical error, this spread represents the emergence of Markovian dissipation; its magnitude and sign are tied to the causal direction of the difference scheme, as analyzed in Appendix~\ref{app:scheme}. Temporal coarse-graining blurs the microscopic coherent oscillations, providing the mechanism through which macroscopic irreversible coefficients, such as local friction ($\gamma$) and kinematic viscosity ($\nu$), physically manifest.

\begin{figure}[htbp]
    \centering
    \includegraphics[width=\columnwidth]{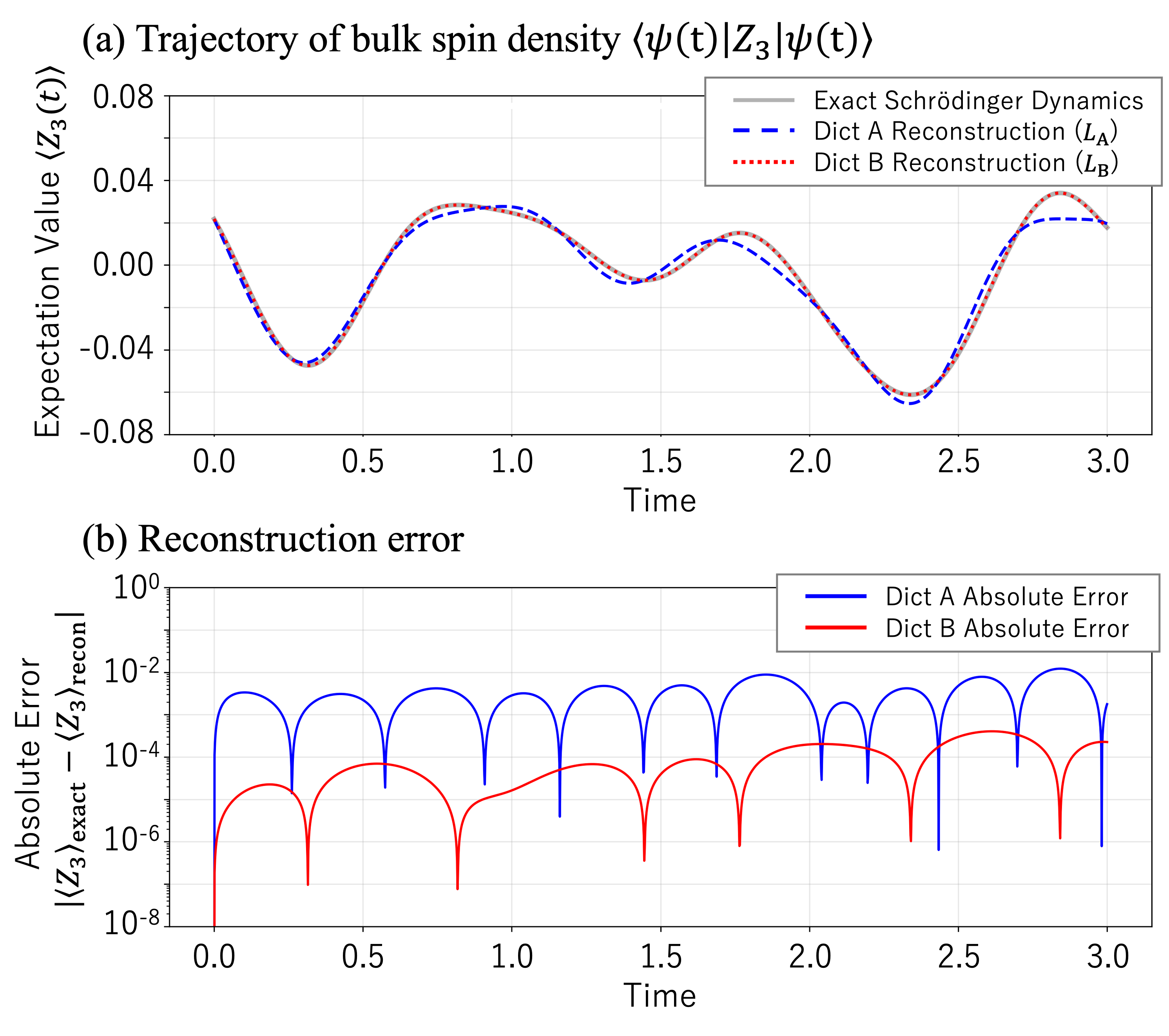}
    \caption{\textbf{Predictive capability.} (a) Trajectory of the macroscopic bulk spin density $\langle Z_3(t) \rangle$ computed via exact Schrödinger dynamics (solid gray line), compared against the analytical reconstructions $e^{L t}\boldsymbol{X}(0)$ using the macroscopic generator $L_{\rm A}$ (Dict A, dashed blue line) and the complete generator $L_{\rm B}$ (Dict B, dotted red line). (b) Absolute reconstruction error for both dictionaries. The complete dictionary (Dict B) captures the Poincaré revivals without artificial damping across 2,000 steps, maintaining an absolute error bounded within $10^{-3}$. Conversely, the macroscopic dictionary (Dict A) deviates, demonstrating the breakdown of the Markovian approximation when unobserved memory effects are neglected.}
    \label{fig:q8_reconstruction}
\end{figure}

The precision of our framework translates directly into predictive capability. In Fig.~\ref{fig:q8_reconstruction}, we evaluate the analytical time evolution of the observables generated by $e^{L t}\boldsymbol{X}(0)$ against the exact Schrödinger dynamics, comparing the performance of the complete dictionary (Dict B) and the macroscopic dictionary (Dict A). As a representative macroscopic variable, we focus on the expectation value of the localized spin density in the bulk (site 3), denoted as $\langle Z_3(t) \rangle$. This bulk observable is ideally suited to visualize both the transient local relaxation and the subsequent quantum revivals caused by boundary reflections.

When employing the complete dictionary (Dict B, $65,535$ observables), the extracted Liouvillian contains virtually no artificial dissipation. Despite learning from a restricted temporal window, the gEDMD reconstruction successfully traces the exact chaotic dynamics and complex Poincaré revivals over an evolution of $2,000$ time steps. The absolute error is constrained within $10^{-3}$. This confirms the algorithmic fidelity of our approach and is consistent with the fact that true macroscopic irreversibility cannot arise without information loss.

Conversely, when relying solely on the macroscopic dictionary (Dict A, $15$ observables), the reconstruction deviates from the exact Schrödinger dynamics, with errors exploding toward $10^{-2}$.
While the macroscopic generator $L_{\rm A}$ manages to qualitatively reproduce the early quantum revivals by utilizing the emergent unstable poles ($\mathrm{Re}(\lambda) > 0$) to express the re-amplification of the observable amplitude, it eventually fails to maintain quantitative accuracy due to the neglected memory effects.
This deviation exposes the limitation of the Markovian approximation when tracking only macroscopic variables without appropriate temporal coarse-graining. The missing information manifests as persistent memory effects from the unobserved environment, motivating our application of the Mori-Zwanzig projection formalism to systematically trace out the bath and extract the effective macroscopic dynamics across different coarse-graining scales.

\section{Results II: Emergence of Dissipation via System-Environment Interaction}

Having established the fundamental dichotomy between exact unitary dynamics and temporal coarse-graining, we now investigate the microscopic origin of macroscopic dissipation. In classical mechanics, friction is introduced phenomenologically; in quantum mechanics, however, true dissipation arises solely from the entanglement and information leakage between a subsystem and its unobserved environment---a process central to decoherence. To visualize this process, we logically partition the quantum spin chain into a target system (S) and its surrounding environment (Env) and employ a quench protocol where the total Hamiltonian $H(t)$ is abruptly switched at $t = t_{\rm q}$:
\begin{equation}
    H(t) = 
    \begin{cases} 
        H_{\text{iso}} = H_{\rm S} + H_{\text{Env}} & (t < t_{\rm q}) \\
        H_{\text{coupled}} = H_{\text{iso}} + V_{\text{int}} & (t \ge t_{\rm q})
    \end{cases}
\end{equation}
where $H_{\rm S}$ and $H_{\text{Env}}$ represent the internal unitary dynamics of the target and environment, respectively. For the $N=20$ system, the boundary interaction $V_{\text{int}}$ consists of the nearest-neighbor (NN) and next-nearest-neighbor (NNN) terms that bridge the partition at site $i=3$:
\begin{equation}
    V_{\text{int}} = J(X_3 X_4 + Y_3 Y_4) + \Delta Z_3 Z_4 + J_2(Z_2 Z_4 + Z_3 Z_5)
\end{equation}
In Phase 1 ($t < t_{\rm q}$), the subsystems evolve independently, ensuring that any subsequent dissipation is purely interaction-induced. In Phase 2 ($t \ge t_{\rm q}$), the abrupt activation of $V_{\text{int}}$ initiates the information flow and subsequent decoherence.

\begin{figure}[htbp]
    \centering
    \includegraphics[width=\columnwidth]{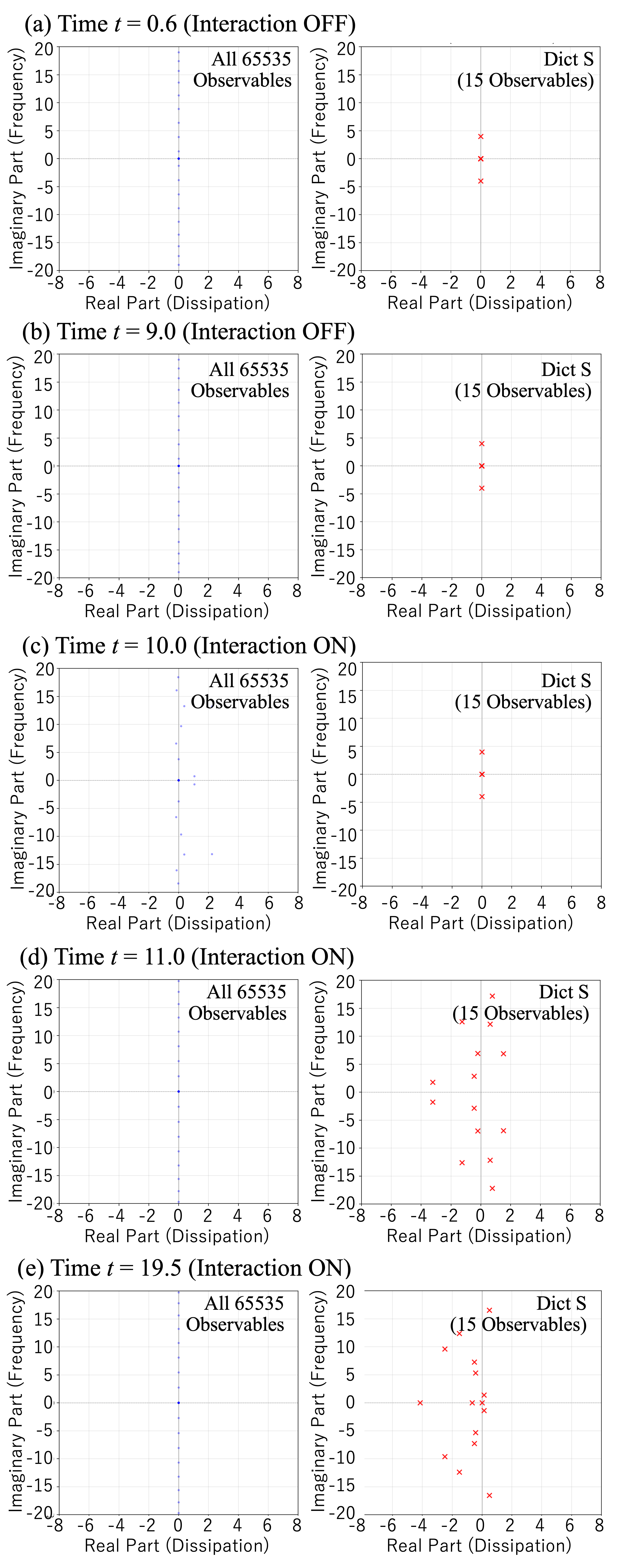}
    \caption{\textbf{Onset of decoherence in an 8-qubit quench ($t_{\rm q} = 10.0$).} The left column shows the spectra obtained using the complete basis of all $65,535$ observables, while the right column shows the spectra extracted using the restricted target dictionary (Dict S). The exact system's eigenvalues remain strictly on the imaginary axis ($\text{Re}(\lambda)=0$), demonstrating absolute unitarity. In contrast, the spectrum of Dict S scatters across the complex plane ($\text{Re}(\lambda) \neq 0$) after contact, visualizing both the rapid loss of local information to the bath (negative real parts) and the subsequent quantum echoes from the finite environment (positive real parts). (The continuous temporal evolution of these spectra is provided as Supplementary Video 2.)}
    \label{fig:q8_contact}
\end{figure}

We first demonstrate this mechanism using the $N=8$ qubit system ($t_{\rm q} = 10.0$), partitioned into a target (S, qubits $0-1$) and an environment (Env, qubits $2-7$). Figure~\ref{fig:q8_contact} compares the complex plane spectra of the complete basis of all $65,535$ observables (representing the entire system) against the restricted target dictionary (Dict S, $15$ observables), which consists of local spin densities ($X_i, Y_i, Z_i$) and all non-trivial two-site correlations ($X_0 X_1, X_0 Y_1, \dots, Z_0 Z_1$) within the two-qubit target region. Before the interaction ($t < 10.0$), both spectra reflect purely unitary dynamics. Upon contact ($t = 10.0$), while the full universe recovers its strict reversibility on the imaginary axis, the eigenvalues of Dict S permanently scatter into the left-half plane. This provides direct numerical evidence that macroscopic dissipation is an emergent consequence of projecting dynamics onto a restricted observable basis. This spatial projection (tracing out the environment) serves as the first necessary condition for irreversibility; however, as we will explore in Section V, stable macroscopic transport further demands appropriate temporal coarse-graining.

\begin{figure}[htbp]
    \centering
    \includegraphics[width=\columnwidth]{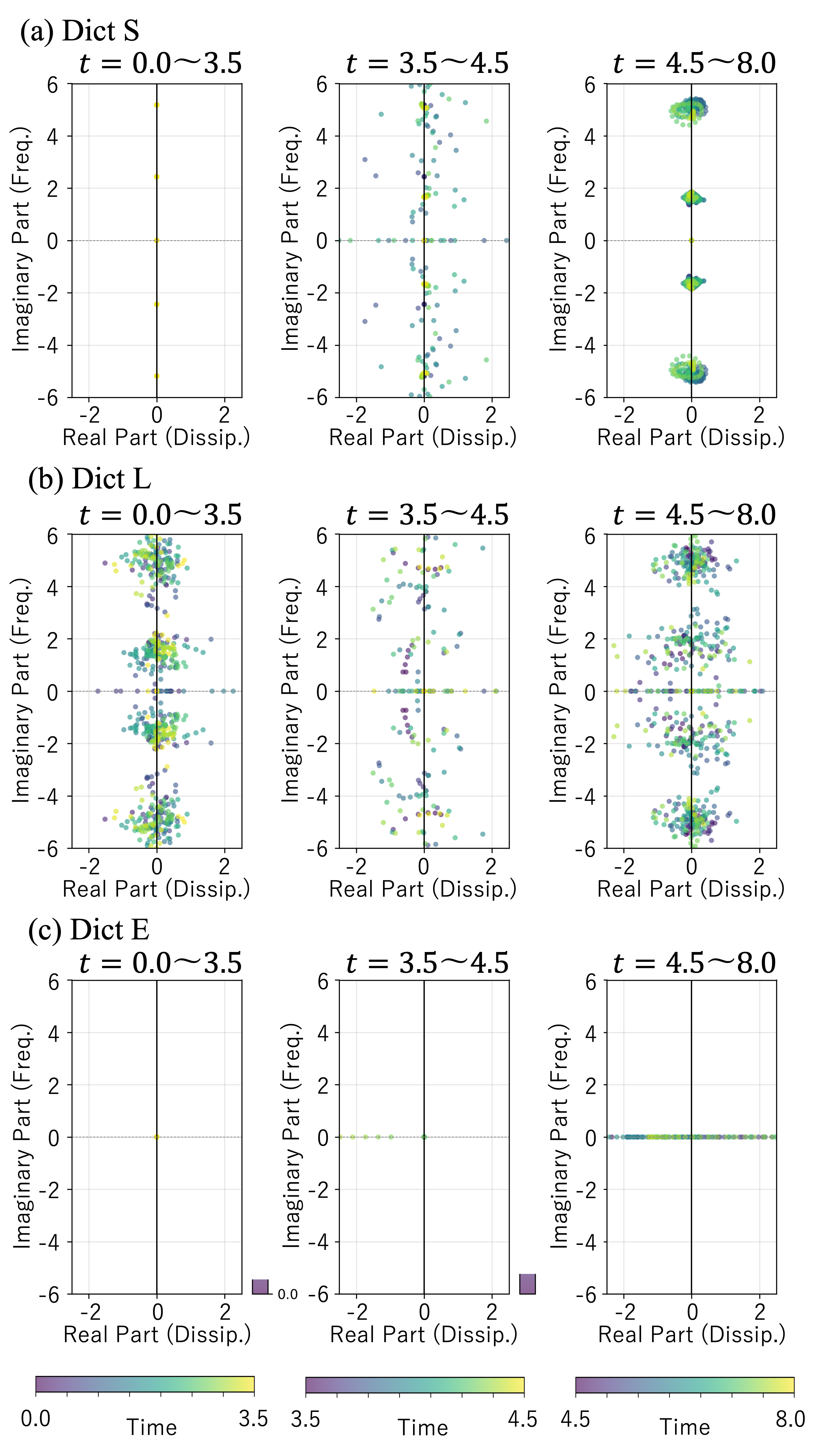}
    \caption{\textbf{Complex plane spectra in the 20-qubit system ($t_{\rm q} = 4.0$).} (a) Dict S (Target) shows localized dissipation as coherence is lost. (b) Dict L (Macroscopic Pointer) exhibits the widest spectral spread across both left and right planes, reflecting its role in mediating the bidirectional flow of information. (c) Dict E (Environment Energy) oscillates along the real axis, reflecting the intermittent energy exchange between S and Env. (An animated visualization of these dynamics is available as Supplementary Video 3.)}
    \label{fig:q20_complex}
\end{figure}

To explore this phenomenon closer to the thermodynamic limit, we analyze the $N=20$ qubit system ($t_{\rm q} = 4.0$), partitioned into a target (S, qubits $0-3$) and an environment (Env, qubits $4-19$). We probe the environment through two distinct functional lenses: Dict L serves as a macroscopic fluid pointer comprising $61$ observables, tracking spin densities ($Z_i$), nearest-neighbor correlations ($Z_i Z_{i+1}$), and spin currents ($X_i Y_{i+1}, Y_i X_{i+1}$); meanwhile, Dict E monitors the total internal energy of the environment as a one-dimensional thermodynamic indicator.

As shown in Fig.~\ref{fig:q20_complex}, Dict L exhibits a broad spectral distribution across both the left ($\text{Re}(\lambda)<0$) and right ($\text{Re}(\lambda)>0$) half-planes. This signifies its role as a dynamic conduit: it absorbs information from S (amplification) and simultaneously leaks it into unobserved microscopic correlations (dissipation). The presence of positive real parts further captures the "quantum echoes" returning from the boundaries, a hallmark of finite-size effects. In contrast, the spectrum of Dict E oscillates along the real axis, indicating the periodic energy "sloshing" between the target and the environment, confirming that the 16-qubit bath is nearing, but has not yet reached, the infinite thermodynamic limit.

\begin{figure}[htbp]
    \centering
    \includegraphics[width=\columnwidth]{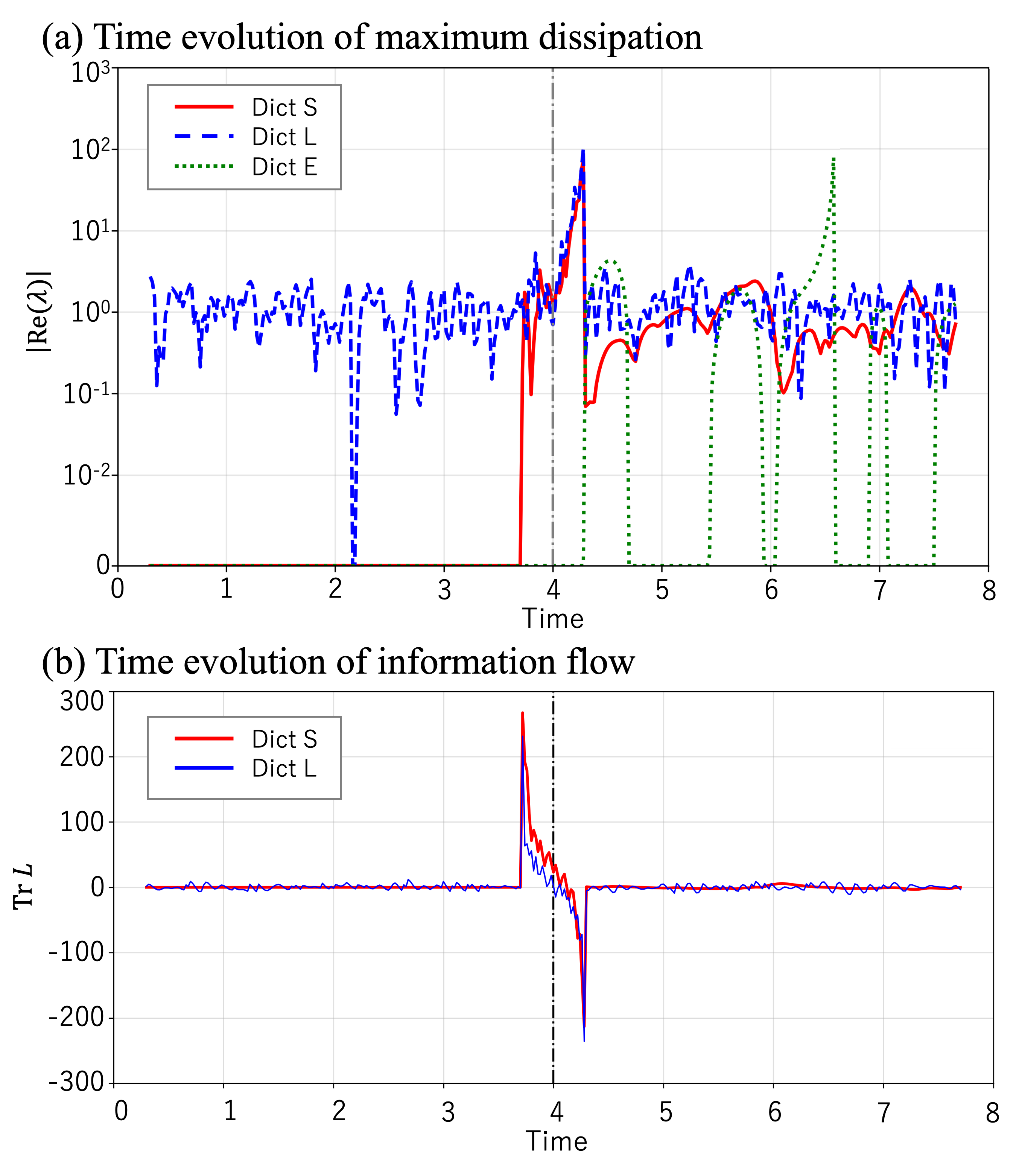}
    \caption{\textbf{Dynamic tracking of dissipation and information flow (20-qubit).} (a) Time evolution of the maximum dissipation pole ($|\text{Re}(\lambda)|$). Dict L (blue) shows constant dissipation even before contact, reflecting the internal thermalization of the chaotic environment, while Dict E (green dotted) oscillates intermittently, reflecting the episodic energy exchange between the target and the environment. (b) The Liouvillian trace ($\sum \text{Re}(\lambda)$) captures the synchronized outflow of information from S (red) and its inflow to L (blue) at the quench, visualizing the decoherence process.}
    \label{fig:q20_time_evolution}
\end{figure}

The dynamical onset of this macroscopic dissipation is quantitatively tracked in Fig.~\ref{fig:q20_time_evolution}. Figure \ref{fig:q20_time_evolution}(a) reveals that prior to contact ($t < 4.0$), Dict L already exhibits a baseline dissipation level ($\sim 10^0$), confirming that the chaotic environment undergoes self-thermalization even when isolated. Upon contact, Dict S undergoes a rapid increase in dissipation as local coherence is lost. This information transfer is quantified through the Liouvillian trace, $\text{Tr}(L) = \sum \text{Re}(\lambda_i)$, which serves as a proxy for phase-space volume contraction. Figure \ref{fig:q20_time_evolution}(b) reveals an abrupt outflow of information (negative trace) from the target system S exactly at the quench, matched by a corresponding influx (positive trace) into the macroscopic pointer L. This synchronized surge provides a data-driven visualization of the "quantum measurement" process: the exact moment when localized quantum information is irreversibly swept away by macroscopic currents and dissipated into the environmental degrees of freedom.

\section{Results III: Temporal Coarse-Graining and the Emergence of Macroscopic Hydrodynamics}

Having visualized the microscopic origins of decoherence via spatial projection, we now focus on the macroscopic phenomenological equations governing the transport of the conserved spin density $Z$. To robustly extract the universal macroscopic bulk properties while preserving the true physical fluctuations, we evaluate the exact-derivative and finite-difference gEDMD using an ensemble of 500 independent random initial states. As discussed in Section II, this large statistical ensemble allows us to naturally average out stochastic noise without relying on artificial superoperator manipulations, thereby avoiding bias in the extracted coefficients.

To accurately capture the macroscopic transport and local thermalization up to length-3 correlations, our generalized dictionary is constructed with a total of 149 observables. Specifically, it comprises local spin densities ($Z_i$, 20 elements), nearest-neighbor spin currents ($X_i Y_{i+1} - Y_i X_{i+1}$, 19 elements), nearest-neighbor longitudinal correlations ($Z_i Z_{i+1}$, 19 elements), symmetric kinetic exchange terms ($X_i Y_{i+1} + Y_i X_{i+1}$, 19 elements), and length-3 composite correlations including three-site densities ($Z_i Z_{i+1} Z_{i+2}$) and current-density interactions (totaling 72 elements).

The matrix elements of the extracted effective Liouvillian, $(L_{\text{open}})_{\alpha, \beta}$, directly dictate the time-evolution of the macroscopic variables. By mapping the row corresponding to the local spin current $J_k$ onto a discrete generalized Navier-Stokes equation, we extract the local hydrodynamic coefficients, namely the mechanical elasticity/pressure gradient ($c_k^2$), the local friction ($\gamma_k$), and the kinematic viscosity ($\nu_k$), using the exact matrix elements:
\begin{align}
    c_k^2 &= \frac{1}{2} \left( (L_{\text{open}})_{J_k, Z_k} - (L_{\text{open}})_{J_k, Z_{k+1}} \right), \\
    \gamma_k &= -(L_{\text{open}})_{J_k, J_k}, \\
    \nu_k &= \frac{1}{2} \left( (L_{\text{open}})_{J_k, J_{k-1}} + (L_{\text{open}})_{J_k, J_{k+1}} \right).
\end{align}
Assuming an overdamped regime where the current relaxes much faster than the density accumulation, the macroscopic diffusion coefficient $D_k$ is defined via the Einstein relation:
\begin{equation}
    D_k = \frac{c_k^2}{\gamma_k}.
\end{equation}
Simultaneously, we evaluate the direct self-relaxation rate of the local spin density, defined as $D_{Z, k} = -(L_{\text{open}})_{Z_k, Z_k} / 2$. Since the total spin is a strictly conserved quantity dictated by the continuity equation ($\dot{Z}_k + \nabla J_k = 0$), $D_{Z, k}$ must theoretically remain zero, providing a strict validity check for our unconstrained data-driven extraction.

\begin{figure}[htbp]
    \centering
    \includegraphics[width=\columnwidth]{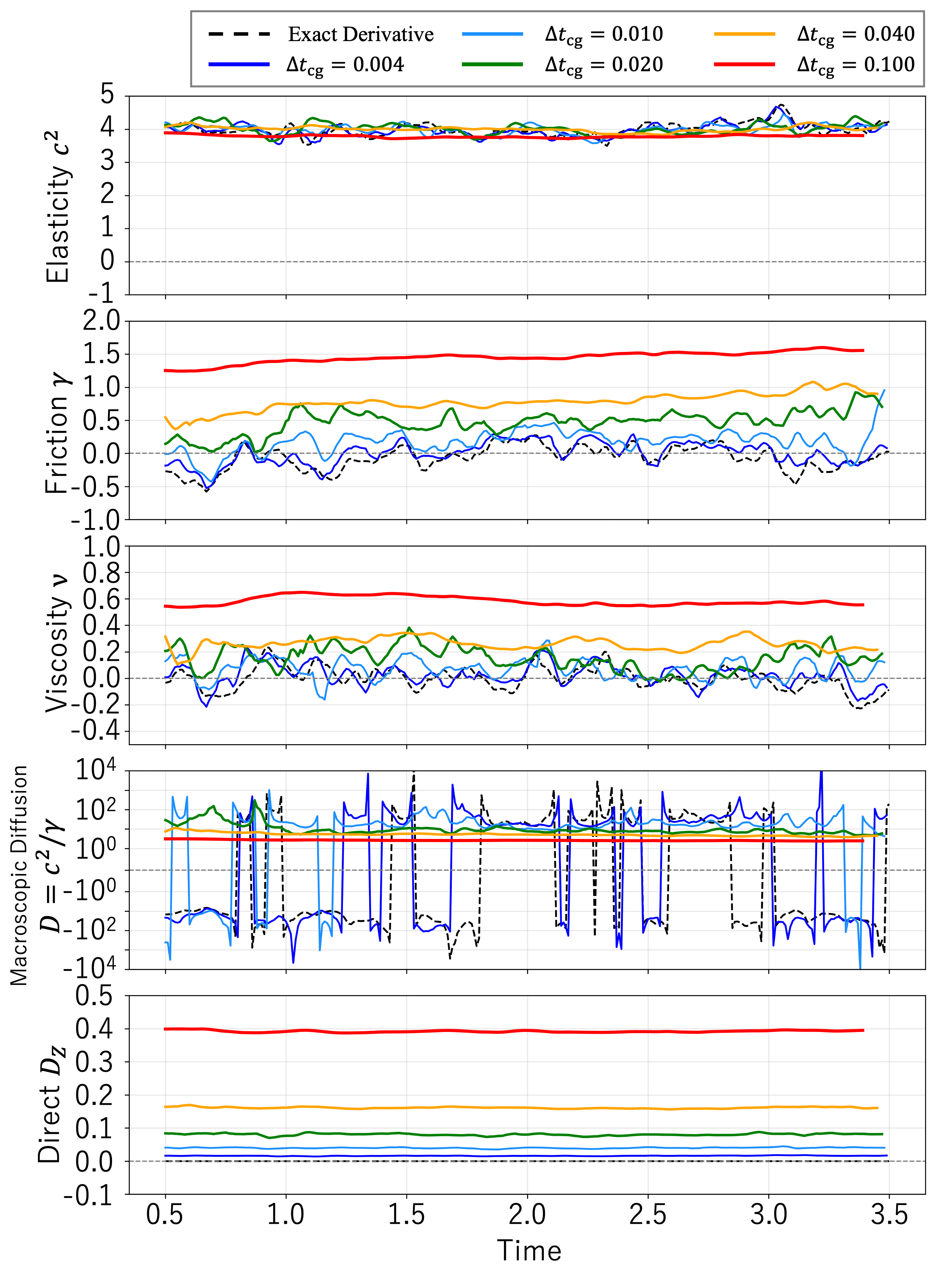}
    \caption{\textbf{Time evolution of hydrodynamic coefficients across varying temporal resolutions.} The coefficients are spatially averaged (median) over the bulk (sites $i = 2 \sim 17$) using a sliding window ($\tau=1.0$). While the exact derivative extracts a robust elasticity $c^2$, both the friction $\gamma$ and kinematic viscosity $\nu$ oscillate around zero. By introducing temporal coarse-graining, the reversible microscopic oscillations are averaged out, allowing strictly positive macroscopic friction $\gamma$ and viscosity $\nu$ to emerge, establishing a stable diffusion $D$.}
    \label{fig:cg_time_comparison}
\end{figure}

\begin{figure}[htbp]
    \centering
    \includegraphics[width=\columnwidth]{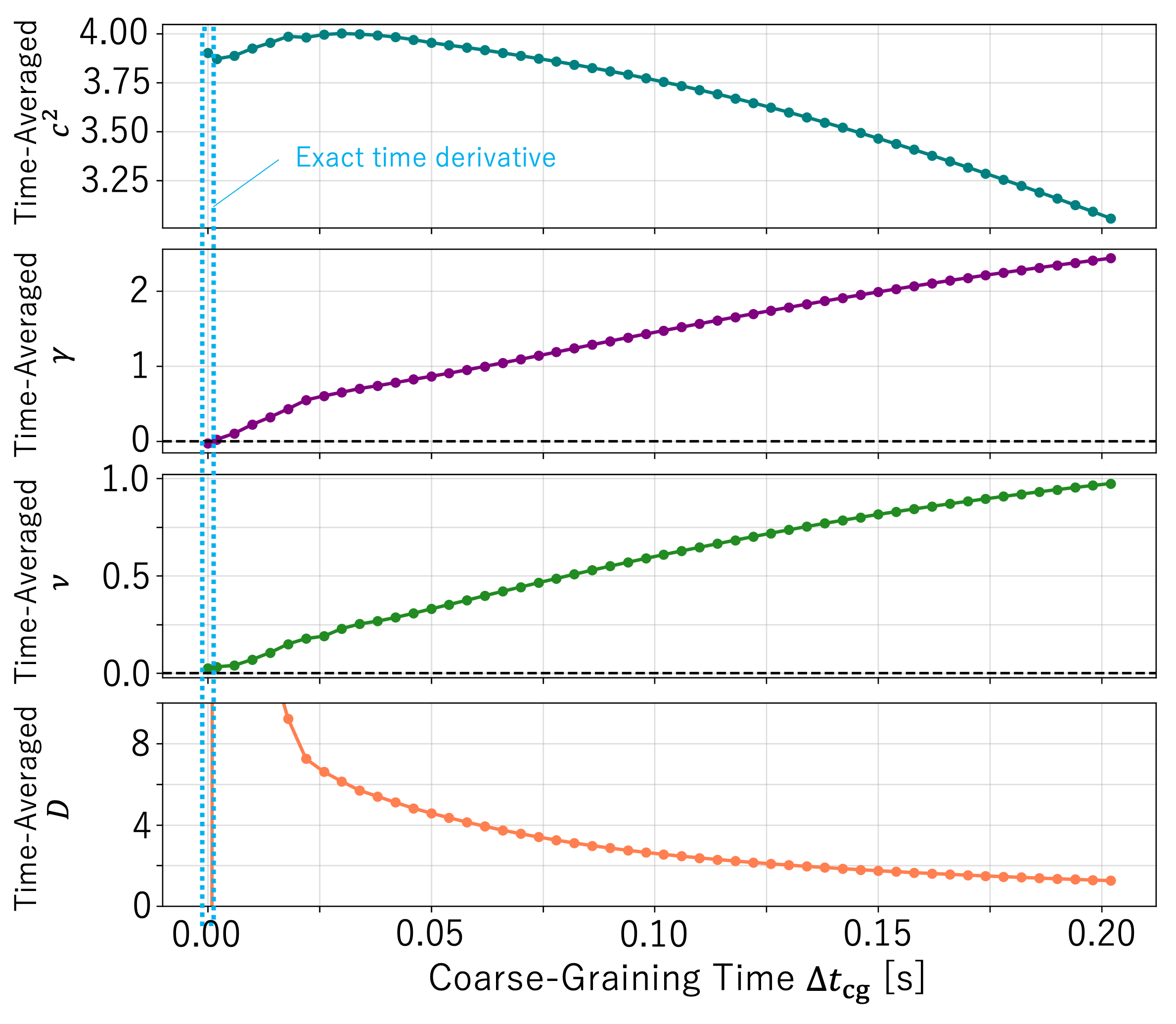}
    \caption{\textbf{Emergence of macroscopic dissipation via temporal coarse-graining.} Time-averaged hydrodynamic coefficients ($c^2, \gamma, \nu$, and $D$) evaluated during the stable hydrodynamic window ($t = 1.0 \sim 2.5$) are plotted against the coarse-graining time $\Delta t_{\rm cg}$. The exact-derivative limit ($\Delta t_{\rm cg} \to 0$) yields zero net dissipation for both friction and viscosity. As $\Delta t_{\rm cg}$ increases, the coefficients exhibit a distinct inflection point around $\Delta t_{\rm cg} \approx 0.025$, marking the characteristic crossover timescale where microscopic quantum coherence is effectively blurred. Following this transition, the irreversible dissipation coefficients ($\gamma, \nu$) show steady monotonic growth, establishing a well-defined and finite macroscopic diffusion regime for $D$. Beyond this intermediate scale, the classical fluid model mathematically collapses due to artificial over-damping (the sinc-filter effect).}    \label{fig:markovian_plateau}
\end{figure}

To reveal the dichotomy between exact unitary mechanics and macroscopic irreversibility, we compare the extracted dynamics across varying temporal coarse-graining scales ($\Delta t_{\rm cg}$). Figure~\ref{fig:cg_time_comparison} illustrates the continuous time evolution of the bulk macroscopic coefficients for different $\Delta t_{\rm cg}$, while Figure~\ref{fig:markovian_plateau} plots the time-averaged values of these coefficients against the degree of coarse-graining during the stable temporal window. 

In the exact-derivative limit ($\Delta t_{\rm cg} \to 0$), the mechanical pressure gradient $c^2$ is accurately captured, maintaining a stable value of approximately 4.0. However, both the local friction $\gamma$ and kinematic viscosity $\nu$ exhibit rapid, reversible fluctuations around zero: within the stable window $t = 1.0 \sim 2.5$, the bulk-median friction averages to $\bar{\gamma} = -0.035$ with a fluctuation amplitude of $0.185$, statistically consistent with zero. This demonstrates that, without temporal blurring, the system preserves strict microscopic reversibility, exhibiting zero net dissipation across all momentum transport channels.

Genuine irreversible hydrodynamics only emerges when we introduce a finite observation timescale. As explicitly captured in the four-panel plot of Figure~\ref{fig:markovian_plateau}, as $\Delta t_{\rm cg}$ increases, the coefficients exhibit a distinct inflection point around $\Delta t_{\rm cg} \approx 0.025$. This marks the characteristic crossover timescale where the microscopic coherent "sloshing" of information is effectively blurred. Following this transition, within an optimal intermediate window ($\Delta t_{\rm cg} \approx 0.04 \sim 0.10$), both spatial and temporal dissipative terms emerge simultaneously. The system mimics a classical fluid with strictly positive friction ($\gamma>0$) and viscosity ($\nu>0$) showing steady monotonic growth, thereby establishing a well-defined and finite macroscopic diffusion regime for $D$. However, if the temporal resolution is excessively degraded (e.g., $\Delta t_{\rm cg} > 0.200$), the coarse-graining width overtakes the macroscopic wave period. This introduces a severe sinc-filter effect that artificially dampens the time derivatives, causing $c^2$ to collapse toward zero and ruining the dynamical equations. The dependence of these results on the choice of the finite-difference scheme, and the distinct role of the causal direction of the inference, are analyzed in Appendix~\ref{app:scheme}.

\begin{figure}[htbp]
    \centering
    \includegraphics[width=\columnwidth]{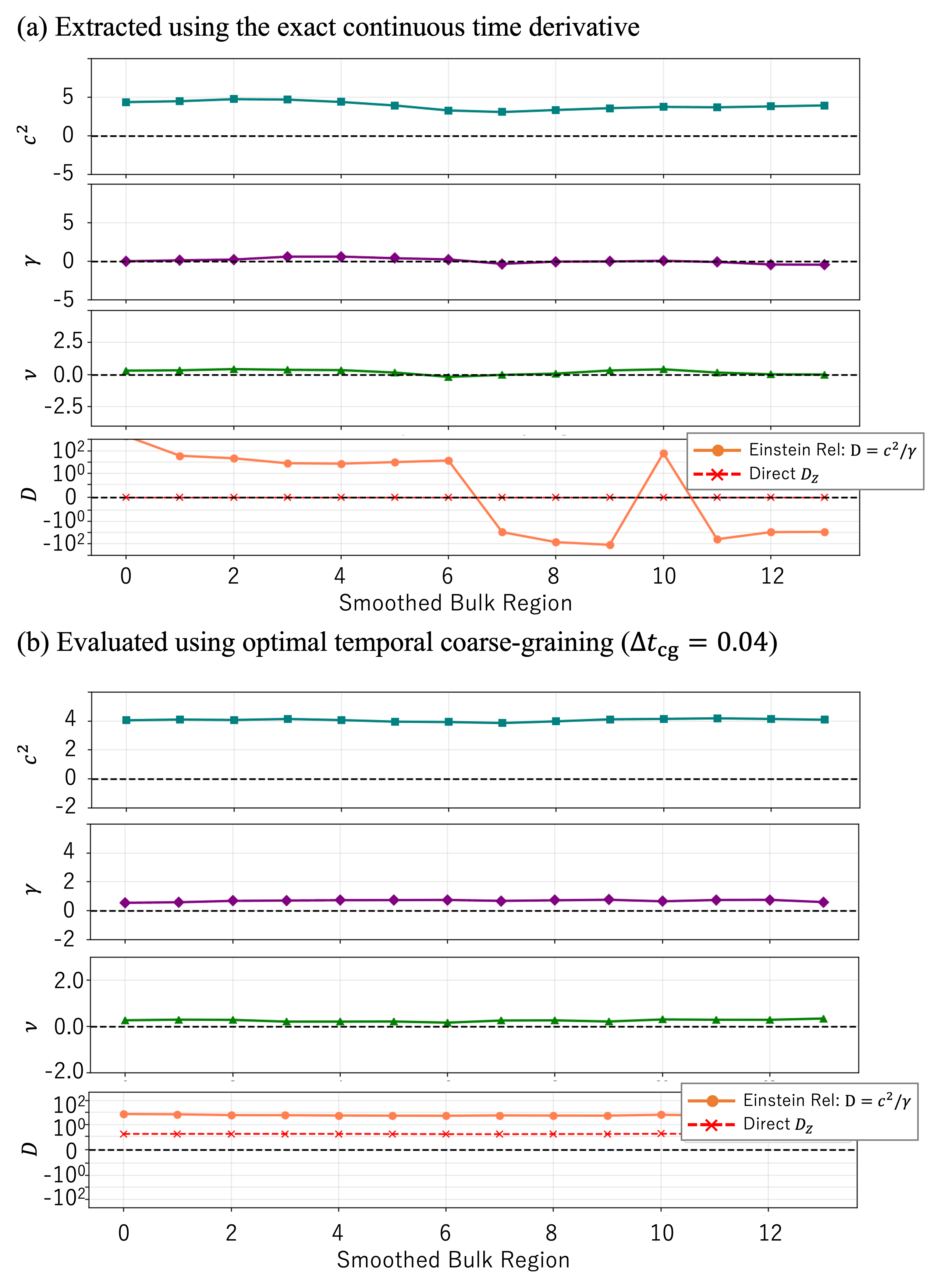}
    \caption{\textbf{Spatial profiles of the emergent hydrodynamics ($t = 0.5 \sim 1.5$).} (a) Extracted using the exact continuous time derivative, the friction $\gamma$ exhibits zero net dissipation, failing to establish a coherent macroscopic diffusion $D$. (b) Evaluated using optimal temporal coarse-graining ($\Delta t_{\rm cg} = 0.04$), the reversible fluctuations are suppressed, revealing a stable, uniformly positive friction $\gamma$ across the bulk and establishing a consistent macroscopic diffusion $D$. The direct density relaxation $D_Z$ (red crosses) remains exactly zero globally in the exact-derivative limit, preserving strict microscopic conservation. Under temporal coarse-graining, $D_Z$ acquires a small positive value, mathematically reflecting the fundamental temporal blurring of the local continuity equation over the finite observation window $\Delta t_{\rm cg}$. (Animated spatial profiles for the exact and coarse-grained dynamics are provided as Supplementary Videos 4 and 5, respectively.)}
    \label{fig:spatial_profiles}
\end{figure}

The impact of this temporal coarse-graining is spatially corroborated in Figure~\ref{fig:spatial_profiles}, which contrasts the spatial profiles of the bulk during the stable timeframe ($t = 0.5 \sim 1.5$). Under the exact derivative (Fig.~\ref{fig:spatial_profiles}(a)), $\gamma$ oscillates across the spatial domain, preventing the formation of a meaningful diffusion coefficient. Conversely, evaluating the identical data under optimal temporal coarse-graining ($\Delta t_{\rm cg} = 0.04$, Fig.~\ref{fig:spatial_profiles}(b)) reveals a uniformly positive friction $\gamma$ across the entire bulk, establishing a well-defined macroscopic diffusion $D$. Notably, under the exact continuous time derivative, the direct self-relaxation rate $D_Z$ remains exactly zero globally, confirming that our unconstrained framework respects strict microscopic conservation. When optimal temporal coarse-graining is applied ($\Delta t_{cg}=0.04$), $D_Z$ acquires a small positive value, mathematically reflecting the fundamental temporal blurring of the local continuity equation over the finite observation window. This demonstrates that the dominant classical diffusion must physically arise through the intermediate generation of current and momentum dissipation, rather than through a direct density decay phenomenological shortcut.

\section{Results IV: Predictive Capability and Dictionary-Size Dependence}

\begin{figure*}[htbp]
    \centering
    \includegraphics[width=\textwidth]{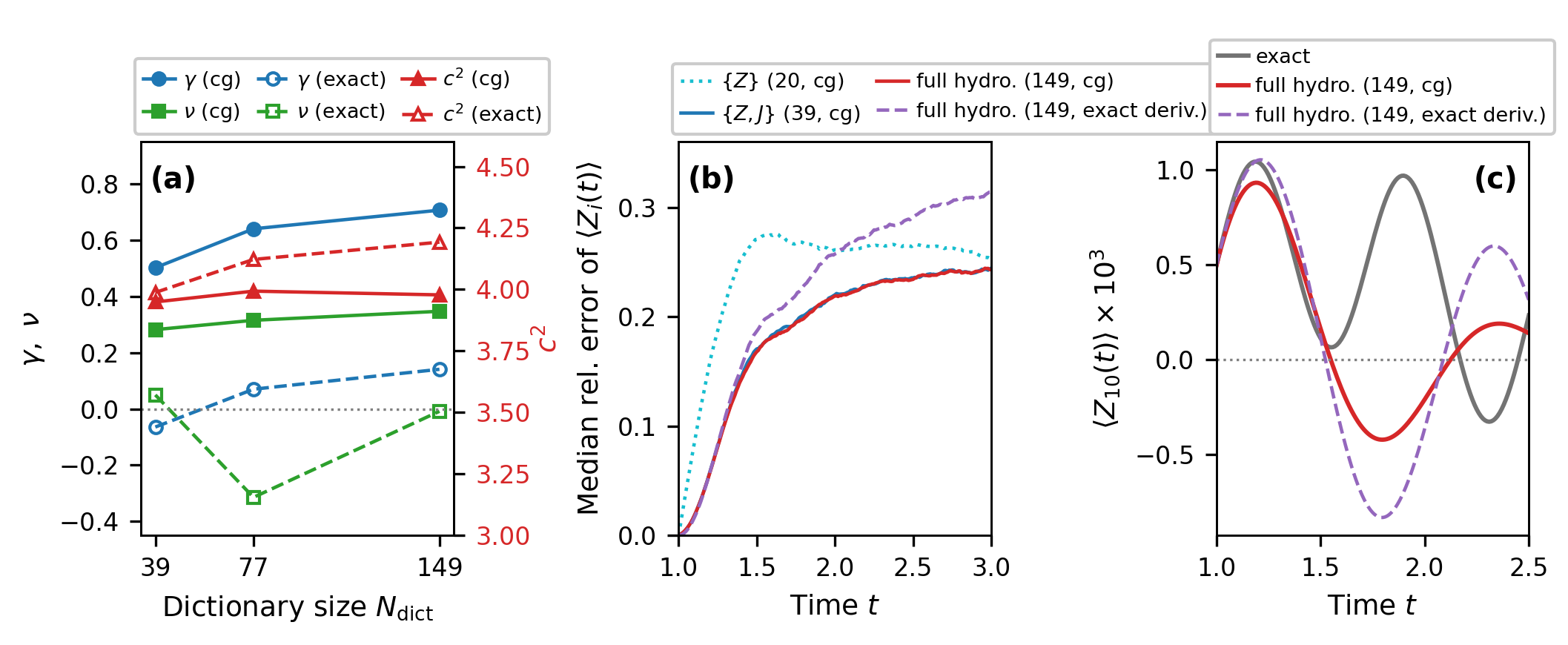}
    \caption{\textbf{Dictionary-size dependence and predictive capability.} (a) Hydrodynamic coefficients extracted from nested sub-dictionaries of sizes $N_{\rm dict} = 39$ ($\{Z_i, J_i\}$), $77$ (adding the length-2 correlations), and $149$ (full hydrodynamic dictionary), using the same 500-trajectory ensemble and training window $t \in [1.0, 2.0]$. Under temporal coarse-graining ($\Delta t_{\rm cg} = 0.04$, filled symbols), the friction $\gamma$ and viscosity $\nu$ converge with successively halving increments, while the exact-derivative values (open symbols) scatter around zero. The elasticity $c^2$ (red triangles, right axis) is insensitive to both the dictionary size and the differentiation paradigm, varying by less than $6\%$ ($3.95$--$4.19$). (b) Median relative prediction error of the local spin densities $\langle Z_i(t) \rangle$ over the 500-trajectory ensemble, for the analytical forward prediction $e^{L(t-t_0)}\boldsymbol{X}(t_0)$ launched at $t_0 = 1.0$. The minimal hydrodynamic dictionary $\{Z, J\}$ ($N_{\rm dict}=39$) is indistinguishable from the full hydrodynamic dictionary, whereas the density-only dictionary ($N_{\rm dict}=20$) is clearly inferior. Since even the 149-element dictionary is a minute projection of the full operator space, the coarse-grained generators saturate at the typicality noise floor, while the exact-derivative generator---lacking the dissipation that would encode the neglected memory---accumulates error monotonically. (c) Representative single-trajectory prediction of the bulk spin density $\langle Z_{10}(t) \rangle$ (full hydrodynamic dictionary). The coarse-grained generator (solid red) quantitatively tracks the exact dynamics over the friction timescale $\sim \gamma^{-1}$ and subsequently relaxes toward the correct infinite-temperature value, whereas the exact-derivative generator (dashed purple) continues to oscillate without relaxation.}
    \label{fig:dict_prediction}
\end{figure*}

Having established how the hydrodynamic coefficients emerge, we now quantify two practical aspects of the framework: how large the observable dictionary must be, and to what extent the extracted generator $L$ can \textit{predict} the subsequent evolution of the macroscopic observables. To this end, we repeat the extraction of Sec.~V on nested sub-dictionaries of the full 149-element hydrodynamic dictionary---the local densities alone ($\{Z_i\}$, $N_{\rm dict}=20$), densities and currents ($\{Z_i, J_i\}$, $N_{\rm dict}=39$), all length-2 observables ($N_{\rm dict}=77$), and the full hydrodynamic dictionary including length-3 correlations ($N_{\rm dict}=149$)---and evaluate the analytical forward prediction $e^{L (t - t_0)} \boldsymbol{X}(t_0)$ against the exact Schr\"odinger dynamics. We emphasize that even this largest dictionary remains a minute projection of the full operator space of the $N=20$ chain (149 out of $4^{20}-1 \approx 1.1\times 10^{12}$ Hermitian operators); it is the complete \textit{hydrodynamic} dictionary, not the complete Pauli basis of Sec.~III, and it therefore traces out essentially all microscopic degrees of freedom.

Figure~\ref{fig:dict_prediction}(a) shows that the extracted coefficients converge rapidly with the dictionary size. Under temporal coarse-graining ($\Delta t_{\rm cg} = 0.04$), the friction grows from $\gamma = 0.50$ ($N_{\rm dict}=39$) to $0.64$ ($77$) and $0.71$ ($149$), with successively halving increments indicating near-convergence at the length-3 level; the viscosity $\nu$ behaves analogously. The elasticity $c^2$ is essentially independent of both the dictionary size and the differentiation paradigm, consistent with its purely mechanical, reversible origin. In the exact-derivative limit, $\gamma$ and $\nu$ remain scattered around zero for every dictionary, confirming that no enlargement of the observation space can, by itself, produce net dissipation.

The predictive capability is quantified in Figs.~\ref{fig:dict_prediction}(b) and (c). Three features stand out. First, the coarse-grained generator is genuinely predictive: launched from a single snapshot $\boldsymbol{X}(t_0)$, it quantitatively tracks the exact evolution of the bulk spin density over the friction timescale $\sim \gamma^{-1}$ and then correctly relaxes toward the infinite-temperature equilibrium value [Fig.~\ref{fig:dict_prediction}(c)]. The residual error saturates at the level set by the typicality fluctuations of the individual random states ($\sim 2^{-N/2}$), which a deterministic reduced model should not---and cannot---reproduce. Second, the exact-derivative generator fails as a predictive model despite being the more faithful representation of the instantaneous dynamics: lacking net dissipation, it continues to oscillate and accumulates error monotonically. Third, and most important for scalability, the prediction quality of the minimal hydrodynamic dictionary $\{Z, J\}$ ($N_{\rm dict} = 39 = 2N-1$) is indistinguishable from that of the full 149-element hydrodynamic dictionary. Reproducing the full microscopic dynamics requires an exponentially large dictionary, as demonstrated in Sec.~III ($4^N - 1$ Pauli strings); predicting the coarse-grained dynamics of the slow, conserved observables requires only $O(N)$ variables. Temporal coarse-graining is precisely what closes this gap, converting the neglected memory effects into the stable dissipative coefficients that render the small dictionary self-contained.

\section{Discussion and Conclusion}

The systematic, data-driven extraction of hydrodynamic coefficients across varying temporal resolutions clarifies how the thermalization of isolated quantum systems should be understood. Historically, a large body of literature has attempted to characterize transport in quantum spin chains by forcing the spin density dynamics into a purely diffusive mold ($\dot{Z} = D_Z \nabla^2 Z$). However, our unconstrained extraction reveals that the direct density-to-density diffusion coefficient is strictly zero ($D_Z = 0$), mathematically preserving the exact microscopic conservation law of the system at all times. The macroscopic dynamics is fundamentally prohibited from taking this phenomenological shortcut; instead, the density gradient must physically generate a mechanical pressure ($c^2$), which drives the spin current ($J$), which in turn is subjected to local friction ($\gamma$) generated by the many-body environment. Bypassing this intermediate momentum generation inevitably leads to a misinterpretation of the underlying timescale and scaling laws.

More importantly, our results expose a fundamental physical dichotomy regarding the origin of this local friction. The exact continuous time derivative inherently preserves the strict unitary nature of the Schrödinger equation, yielding zero net macroscopic dissipation. True macroscopic friction ($\gamma > 0$) and kinematic viscosity ($\nu > 0$) are not absolute microscopic properties; rather, they are emergent phenomenological consequences of the observer's temporal resolution. The scheme analysis of Appendix~\ref{app:scheme} sharpens this statement considerably: the time-symmetric central-difference estimator reproduces the exact-derivative result---zero net friction---at \textit{any} coarse-graining scale, while the backward difference yields the anti-dissipative mirror image of the forward-difference result. A finite temporal resolution alone is therefore not sufficient; positive macroscopic friction additionally requires a causal, forward-in-time direction of the coarse-grained inference, the same directedness that underlies the Onsager regression of fluctuations. In this sense, the arrow of time enters the macroscopic description through the act of forward prediction itself. By introducing temporal coarse-graining, the microscopic coherent oscillations are averaged out, passing through a distinct crossover timescale to establish an intermediate functional regime where the classical Navier-Stokes fluid picture is successfully realized. Crucially, the fact that the extracted coefficients inherently depend on the coarse-graining time $\Delta t_{\rm cg}$ (as evidenced in Fig. 9) does not imply a loss of physical objectivity, but rather highlights the fundamental role of timescale separation in statistical mechanics. In genuine macroscopic thermodynamic systems, an astronomical separation exists between microscopic collision times and macroscopic relaxation times. This immense gap renders this intermediate functional regime virtually infinite and flat, allowing everyday observation instruments to extract seemingly observer-independent, constant friction coefficients. In our finite quantum system, however, this scale separation is exceptionally tight. Consequently, the fluid model only survives within a narrow temporal sweet spot, delicately bounded by zero-friction reversibility at the exact-derivative limit ($\Delta t_{\rm cg} \to 0$) and artificial over-damping due to the sinc-filter effect at larger timescales. Our data-driven framework thus uniquely visualizes the exact physical crossover where absolute reversible mechanics gives way to emergent macroscopic hydrodynamics.

Finally, we comment on the practical scalability of the framework. The gEDMD regression requires as input only the expectation-value time series $\boldsymbol{X}(t)$ and, for the exact-derivative variant, the commutator expectation values $i\langle [H, \boldsymbol{O}] \rangle$; since both $\boldsymbol{O}$ and $[H, \boldsymbol{O}]$ are local operators, the framework never requires access to the full many-body state itself. The exact state-vector propagation employed here ($N = 20$, Hilbert dimension $2^{20}$) serves as a benchmark free of truncation error, but any method capable of delivering accurate local expectation values---tensor-network time evolution, Krylov-subspace or typicality-based propagation, or direct measurements on a quantum simulator---can equally serve as the data source, limited only by the accuracy horizon of that method. The dictionary itself scales benignly: restricting the observables to strings of length at most $k$ yields $O(N \cdot 3^k)$ elements, linear in the system size, and the regression cost is governed by the covariance matrix inversion, $O(N_{\rm dict}^3)$, which is negligible. As demonstrated in Sec.~VI, this small dictionary is sufficient for the coarse-grained prediction of the slow observables, whereas reproducing the full microscopic dynamics would require an exponentially large basis; the temporal coarse-graining is precisely what makes the linear-in-$N$ description self-contained. We also note that the finite-difference paradigm requires no commutator evaluation at all, so the framework is directly applicable to experimental time-series data, for which the exact derivative is unavailable as a matter of principle.

In conclusion, we have introduced a comprehensive framework combining Exact-Derivative and finite-difference generalized Extended Dynamic Mode Decomposition (gEDMD) integrated with the Mori-Zwanzig projection. By deploying a robust ensemble of 500 independent random initial states and constructing a comprehensive dictionary of 149 observables, capturing up to length-3 composite correlations, this completely data-driven framework allowed us to robustly extract the Navier-Stokes hydrodynamic coefficients from the exact unitary dynamics of a chaotic XXZ spin chain. By explicitly expanding the observable dictionary to separate the spin density and the spin current, we visualized the emergence of a genuine macroscopic fluid regime characterized by positive friction and viscosity. We further demonstrated that the resulting coarse-grained generator constitutes a predictive reduced model, forecasting the relaxation of the slow observables toward their thermal values with only $O(N)$ macroscopic variables, and that the emergent friction is inherited from the causal, forward-in-time direction of the temporal coarse-graining, with time-symmetric estimators yielding zero net dissipation at any scale.

\section*{Acknowledgments}
The author gratefully acknowledges the computational resources provided by the NIFS Plasma Simulator (Project ID: NIFS25KIST066) and the IFERC-CSC supercomputer system, where the extensive numerical simulations for this work were performed.
Additionally, the author acknowledges the use of the generative AI models Gemini (Google) and Claude Code (Anthropic) for facilitating discussions on the research concepts, assisting with the numerical analysis and the preparation of the figures, and providing structural organization and English language editing of the manuscript.

\section*{Code availability}
The code that reproduces the results and figures of this work will be made available upon publication at \url{https://github.com/saitos-lab/gedmd-xxz-hydrodynamics}.

\appendix

\section{Dependence of the extracted coefficients on the finite-difference scheme}
\label{app:scheme}

\begin{figure*}[htbp]
    \centering
    \includegraphics[width=\textwidth]{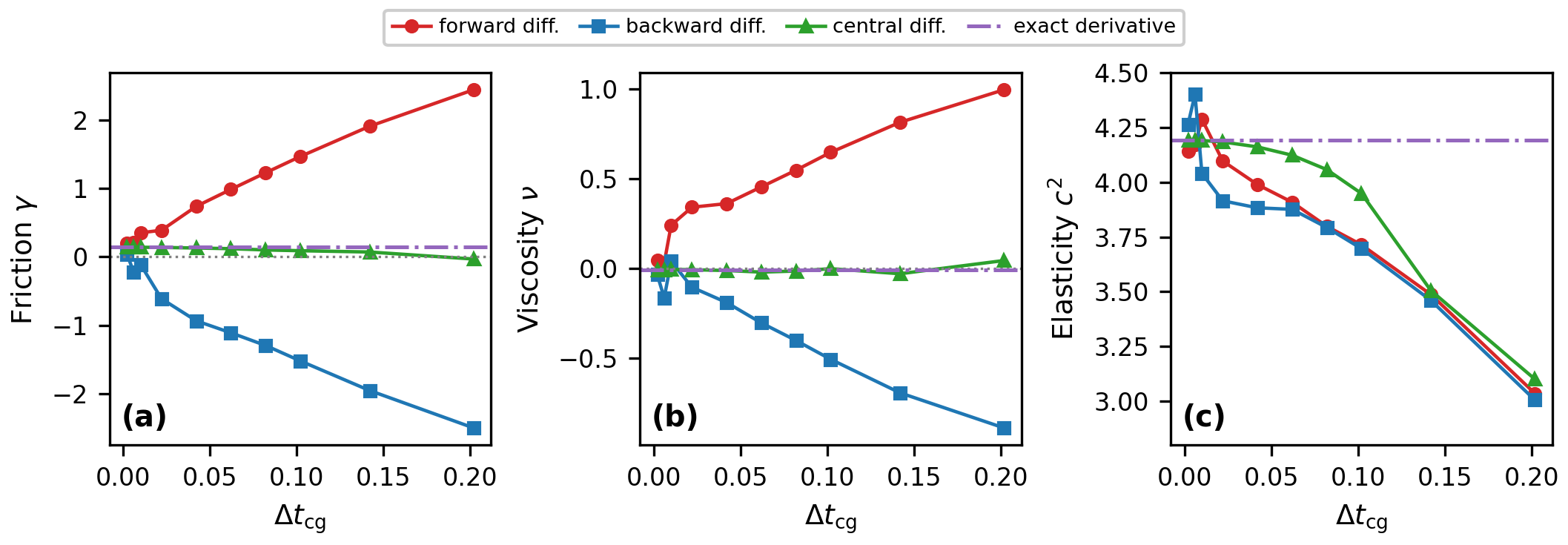}
    \caption{\textbf{Finite-difference scheme dependence of the extracted hydrodynamic coefficients.} (a) Friction $\gamma$, (b) viscosity $\nu$, and (c) elasticity $c^2$ as functions of the coarse-graining time $\Delta t_{\rm cg}$, extracted from the identical 500-trajectory ensemble and training window ($t \in [1.0, 2.0]$) using the forward difference (red circles), backward difference (blue squares), and central difference (green triangles). The horizontal dash-dotted lines mark the exact-derivative values. The backward difference is the anti-dissipative mirror image of the forward difference ($\gamma_{\rm bwd} \approx -\gamma_{\rm fwd}$), while the time-symmetric central difference tracks the exact-derivative result for the friction $\gamma$ and viscosity $\nu$ at every $\Delta t_{\rm cg}$; the elasticity $c^2$, by contrast, is suppressed by the sinc factor common to all three schemes.}
    \label{fig:app_scheme}
\end{figure*}

\begin{figure}[htbp]
    \centering
    \includegraphics[width=\columnwidth]{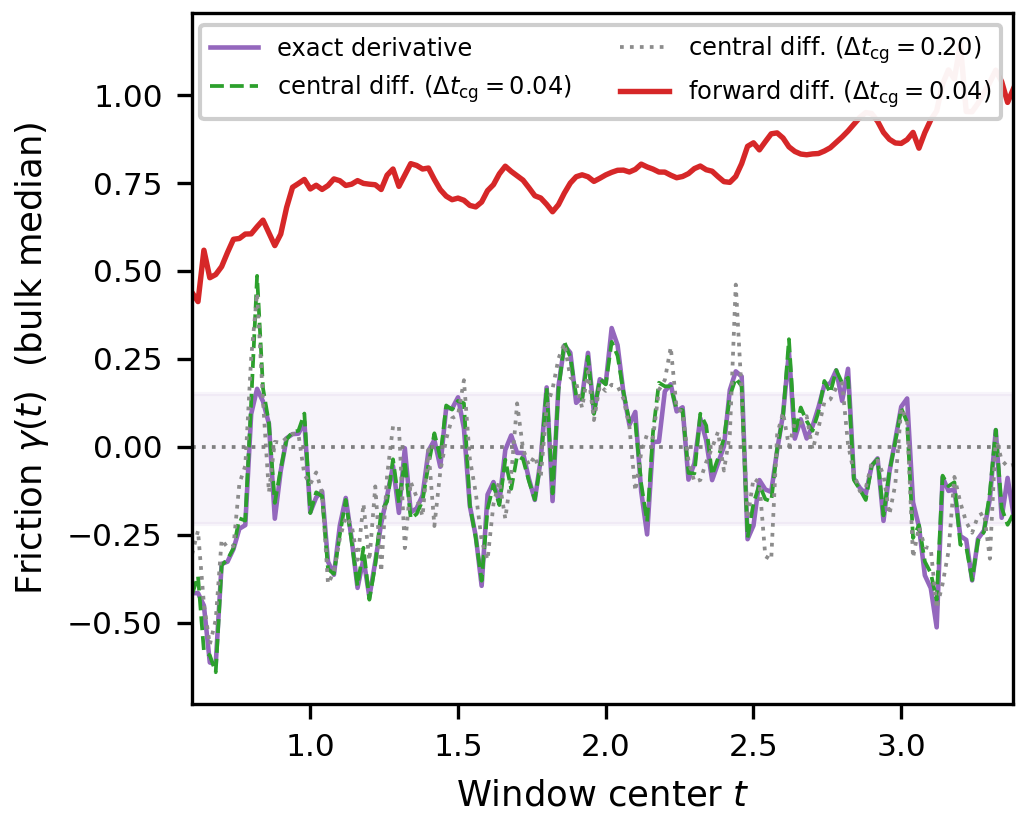}
    \caption{\textbf{Time-resolved friction $\gamma(t)$ for the different estimators} (bulk median, sliding window $\tau = 1.0$, 500-trajectory ensemble). The exact-derivative estimator (purple) and the central differences at $\Delta t_{\rm cg} = 0.04$ (green dashed) and $0.20$ (gray dotted) all oscillate around zero (shaded band: mean $\pm$ one standard deviation of the exact-derivative trace), whereas the forward difference at $\Delta t_{\rm cg} = 0.04$ (red) yields a stable, strictly positive friction.}
    \label{fig:app_timeresolved}
\end{figure}

In the main text, the temporal coarse-graining is implemented as the causal forward difference. Here we quantify how the extracted coefficients depend on this choice by repeating the identical regression with the backward difference, $[\boldsymbol{X}(t) - \boldsymbol{X}(t-\Delta t_{\rm cg})]/\Delta t_{\rm cg}$, and the time-symmetric central difference, $[\boldsymbol{X}(t+\Delta t_{\rm cg}) - \boldsymbol{X}(t-\Delta t_{\rm cg})]/(2\Delta t_{\rm cg})$, all regressed against $\boldsymbol{X}(t)$.

The expected behavior can be read off from a single reversible mode, $X(t) \propto e^{i\omega t}$. The three estimators return effective eigenvalues whose real parts are $(\cos\omega\Delta t_{\rm cg} - 1)/\Delta t_{\rm cg} = -\omega^2 \Delta t_{\rm cg}/2 + O(\Delta t_{\rm cg}^3)$ for the forward difference, $+\omega^2 \Delta t_{\rm cg}/2 + O(\Delta t_{\rm cg}^3)$ for the backward difference, and exactly zero for the central difference, since $\sin(\omega\Delta t_{\rm cg})/\Delta t_{\rm cg}$ is purely real. A time-symmetric estimator therefore cannot generate net dissipation from reversible data at any coarse-graining scale, whereas the forward and backward differences acquire apparent damping and amplification of opposite signs, growing linearly in $\Delta t_{\rm cg}$ and quadratically in the mode frequency---which is precisely why the fast microscopic oscillations, rather than the slow hydrodynamic modes, are converted into friction. The imaginary parts, however, are identical for all three schemes, $\mathrm{Im}(\lambda) = \sin(\omega\Delta t_{\rm cg})/\Delta t_{\rm cg} = \omega\,\mathrm{sinc}(\omega\Delta t_{\rm cg})$, uniformly suppressed below the true frequency $\omega$; consequently the elasticity $c^2$, which derives from this reactive part of the spectrum, is degraded by the same sinc factor in every scheme, since the time symmetry of the central difference protects the dissipative, time-antisymmetric coefficients $\gamma$ and $\nu$ but not the reactive, time-symmetric $c^2$---the frequency reduction being itself a time-symmetric error.

Figure~\ref{fig:app_scheme} confirms this picture for the actual 20-qubit data. The backward-difference friction and viscosity mirror the forward-difference results, $\gamma_{\rm bwd} \approx -\gamma_{\rm fwd}$ and $\nu_{\rm bwd} \approx -\nu_{\rm fwd}$, over the entire range of $\Delta t_{\rm cg}$; this anti-dissipative branch is the time-reversed counterpart of the emergent friction, as required by the reversibility of the underlying dynamics. The central-difference values remain close to the exact-derivative reference throughout. The elasticity $c^2$, in contrast, is common to all three difference schemes, degrading only through the sinc-filter suppression at large $\Delta t_{\rm cg}$.

The time-resolved analysis of Fig.~\ref{fig:app_timeresolved} makes the statistical statement precise. Within the stable window $t = 1.0 \sim 2.5$, the bulk-median friction of the exact-derivative estimator averages to $\bar{\gamma} = -0.035$ with a standard deviation of $0.185$, and the central-difference estimator at $\Delta t_{\rm cg} = 0.04$ is nearly indistinguishable from it, both point-wise in time and statistically ($\bar{\gamma} = -0.035 \pm 0.183$); even at $\Delta t_{\rm cg} = 0.20$ the central difference remains consistent with zero ($\bar{\gamma} = -0.022 \pm 0.177$). The forward difference at $\Delta t_{\rm cg} = 0.04$, by contrast, yields $\bar{\gamma} = +0.759$ with a fluctuation of only $0.037$. The emergent friction reported in the main text is thus entirely carried by the time-directed component of the coarse-grained inference.

The collapse of $c^2$ at large $\Delta t_{\rm cg}$ is thus a property of the finite-difference estimator itself, not a physical effect: it stems from the sinc suppression of the reactive part of the spectrum and vanishes in the exact-derivative limit $\Delta t_{\rm cg}\to 0$. We employ the forward difference in the main text because it is the operational estimator of an observer with finite temporal resolution who predicts the future from the present---the physical situation whose consequences this work sets out to isolate---and because, as shown above, its dissipative component isolates the causal, forward-in-time contribution to the emergent friction.

\bibliographystyle{unsrt}

\bibliography{references}

\end{document}